# Infrared light detection using a whispering-gallery-mode optical microcavity

Jiangang Zhu[1,a)], Sahin Kaya Ozdemir[1,b)], Lan Yang[1,c)]

[1]*Department of Electrical and Systems Engineering, Washington University, St. Louis, Missouri 63117, USA*

We demonstrate a thermal infrared (IR) detector based on an ultra-high-quality-factor (Q) whispering-gallery-mode (WGM) microtoroidal silica resonator, and investigate its performance to detect IR radiation at 10 µm wavelength. The bandwidth and the sensitivity of the detector are dependent on the power of a probe laser and the detuning between the probe laser and the resonance frequency of the resonator. The microtoroid IR sensor achieved a noise-equivalent-power (NEP) of 7.46 nW, corresponding to IR intensity of 0.095 mW/cm$^2$.

Infrared (IR) radiation detectors and imaging systems play crucial roles in both military and civil applications as they can enable detection of objects and produce images in the darkest of nights, and in fog or smoke covered areas. They help to build night vision systems, unmanned autonomous vehicles, surveillance systems, and are widely used in firefighting, building maintenance, security, non-contact thermometry, industrial processing and medical diagnostics. IR detectors are broadly classified into two types: quantum – such as photoconductive and photovoltaic, or thermal – such as bolometers and thermopiles[1-4]. Quantum IR detectors are based on photon detection using semiconductor materials with narrow bandgaps or semiconductor-metal structures with small energy barriers. They offer fast response time; however, they require cryogenic cooling at wavelengths longer than a few micrometers to minimize thermally induced transitions and thermal noises. Cooling requirements increase the cost, size, weight and complexity of quantum IR detectors. Thermal IR detectors, on the other hand, are based on measuring the changes in mechanical, electrical, or optical properties of the detector element due to the thermal energy generated upon absorption of IR radiation. They are broadband and can be operated at or above room temperature without cooling. However, they, in general, have lower detectivity and slower response time because warming up or cooling down of detector elements in response to changes in the IR radiation is a relatively slow process.

---
a) Author to whom correspondence should be addressed. Electronic mail: jzhu@seas.wustl.edu.
b) Author to whom correspondence should be addressed. Electronic mail: ozdemir@seas.wustl.edu.
c) Author to whom correspondence should be addressed. Electronic mail: yang@seas.wustl.edu.

An uncooled detector with the performance of a cooled detector would significantly reduce cost, size, weight, and power requirements[4]. Here, we propose to use high-quality factor (Q) whispering gallery mode (WGM) microtoroid resonators as thermal detectors of IR radiation. WGM microresonators[5] have found applications in various fields due to their ultra-high-Q and microscale mode volumes. In particular, they have enjoyed an increasing attention in the field of optical sensing[6-8], providing ultra-high performance quantified by their unprecedented sensitivity and resolution thanks to ultra-narrow linewidth of their resonances. Among many WGM resonators of different shapes and materials, on-chip silica microtoroids on silicon pillars fabricated using standard photolithography and wafer processing techniques have emerged as the platform with the highest sensitivity and are anticipated to achieve single molecule detection, that is yet to be demonstrated. Nanoparticle detection and size measurement at single particle resolution, ultra-low threshold microlasers, and strong-light matter interactions for cavity-QED and optomechanics have been demonstrated, and the thermal behavior and responsivity to thermally induced changes in the proximity of resonator have been well-studied[6, 7, 9-13]. It's possible to improve the thermal sensitivity of a silica microtoroid by tailoring the size of its supporting silicon pillars to enhance the thermal isolation of the resonator structure[14, 15]. Combined with the high absorption of mid- and far-infrared light in silica, these features suggest that WGM silica microtoroid resonators can be tailored to be used as thermal detectors for IR radiation, providing a microscale, resonantly enhanced uncooled thermal IR detector.

In this Letter, we characterize the performance of a silica microtoroid resonator for detecting IR radiation in 10 μm band. The underlying principle of IR detection is as follows. The incident IR radiation absorbed by silica microresonator increases the resonator temperature. As a result the thermo-optic effect kicks in and leads to a change in the refractive index, which in turn shifts the resonance wavelength. The higher the incident IR radiation is, the larger the shift in the resonance wavelength is. The resonance shift can be detected either by scanning the wavelength of a probe light from a tunable laser around a WGM resonance and recording the change in the resonance after each scanning or by locking the wavelength of the probe on the resonance slope and monitoring the intensity fluctuations of the transmitted light.

The experimental setup used to characterize the performance of our IR detector is shown in Fig. 1. We used a pulse-width modulated $CO_2$ laser with a wavelength of 10μm and a maximum output power of 30 W as the IR source to conduct this study, because silica has a strong absorption at 10 μm wavelength. The beam from the $CO_2$ laser was first expanded by a ZnSe lens and then passed through a small pinhole with a diameter of 381 μm to control the illumination area on the microtoroid, and to block the light incident on the surrounding area of the resonator. Only a small fraction of total IR radiation passes through the pinhole and reaches at the resonator. The microtoroid was placed at an angle of 45 degrees on a heat sink to allow optical imaging and to stabilize its base temperature. We used a probe light from a tunable external cavity laser in the 1450 nm band to monitor the effect of the IR radiation on the transmission spectra and the resonance frequency of the microtoroid. The $CO_2$ laser was modulated with an external gating source with a known frequency. As a result, when the probe laser was locked on a resonance slope (Fig. 1b), the transmission contained the frequency component of modulation, which allowed an easy detection with high signal-noise-ratio using a lock-in amplifier (Fig. 1b).

The microtoroid used in the experiments had a major diameter of about 100 μm, and its pillar was etched to have a size less than 2 μm to maximize thermal isolation from the silicon wafer which acts as a thermal sink (Fig. 1c). The fiber taper that was used to couple the probe light in and out of the WGM was in contact with the microtoroid to stabilize the coupling condition and reduce vibration of the fiber taper. In this situation, we measured the loaded Q factor as $7.8 \times 10^6$. Figure 2a shows the transmission spectrum of the system. First, we modulated the IR source at 100 Hz and monitored the transmission of the probe light through fiber-resonator system. Figure 2b shows the frequency spectrum of the measured transmission. Despite the noise components, a prominent peak at 100 Hz is clearly seen above the noise floor, confirming the response of the resonator to IR radiation. The signal-to-noise ratio (SNR) was about 260 in this measurement. The low frequency noise (<200Hz) was mainly from mechanical vibrations of the resonator system and the tunable laser. The narrow distinct peaks (e.g. at 60Hz and 300Hz) are electrical noises. We observed that even the fan noise from instruments was significant enough to disturb the measurement, suggesting that in a real application the resonator-taper system should be well-isolated, probably in a vacuum chamber, to lower the noise floor and increase the SNR.

To study the frequency response of the IR detector, we varied the modulation frequency of the $CO_2$ laser. As shown in Fig. 3, our IR detector shows typical features of frequency response of a low pass RC filter. When the modulation frequency is low, the temperature changes of the microtoroid can keep up with the gated IR radiation and thus the transmission shows square waves with exponentially decaying edges. As the frequency increased, the amplitude of the response decreased and transmission transformed into semi-triangle waves because of the limited bandwidth set by the thermal relaxation time of the resonator.

Next, we studied the response of the IR detector at different regimes of frequency detuning between the probe laser and the cavity resonance. At low powers, the probe laser doesn't have any significant effect on the thermal characteristics of the IR detecting mechanism. Therefore the IR response is the same regardless of whether the probe laser is on the red- or blue- detuned side of the resonance (Fig. 4a). However, when the power of the probe laser is higher than a few µW, there is a noticeable heating effect in the resonator due to its high Q and good thermal isolation. As a result increased temperature in the resonator red-shifts the resonance and affects the IR response differently on the blue- and red- detuned slopes of the resonance[14].

As shown in Fig. 4b, when the probe laser is on the blue-detuned resonance slope, thermal heating by probe laser increases the device temperature and on-resonance thermal locking effect[14,15] tries to stabilize the device temperature. As a result the detector becomes less sensitive to the thermal heating effect of IR radiation. On the other hand, when the probe laser is on the red-detuned slope, the thermal heating effect by the probe laser keeps the resonator temperature in a slightly unstable regime (opposite to thermal locking effect), which allows higher IR response. As shown in Fig. 4b, the IR response is no longer symmetric on both sides of the resonance peak when probe laser power is sufficiently high.

To show the bandwidth of the microtoroid IR detector, we plot the relation between response amplitude and IR modulation frequency (Fig. 5). The frequency response features a passband and a first order drop-off edge. The frequency response measurement allows us to optimally select the modulation frequency of the incident IR radiation. Ideally one would like to keep the modulation frequency in the passband. We found a good agreement between the measured and theoretically predicted cut-off frequencies. For example, for the black curve in Fig. 5, the experimental data gives a cutoff frequency of 80 Hz, and the thermal relaxation time constant of 1.84 ms measured[16] for the same microtoroid corresponds to a lowpass cutoff frequency of 88.4 Hz.

We observed that the frequency detuning and the power of the probe laser not only modified the thermal sensitivity of the IR detector but also affected its bandwidth. When a probe laser is blue-detuned from the resonance frequency, it can induce thermal locking effect if it has sufficient power, and increase the detector bandwidth but results in relatively smaller response-amplitude. This is because thermal locking tend to keep the system around the locking temperature, and will force it to return back to this temperature more rapidly when it drifts away. On the other hand, a red-detuned probe laser can increase the response amplitude in the pass-band but with a reduction in bandwidth. This unique feature allows us to tune the response of the IR detector without modifying its physical structure (Fig. 5). By adjusting the operating power or Q factor of the mode, and detuning of the probe laser, we can tune the sensitivity and bandwidth of the IR detector and optimize its performance at a fixed IR modulation (chopping) frequency.

Finally, we estimated the noise equivalent power (NEP) of our IR detector. Note that the measurements were performed in normal room conditions without actively stabilizing the sensor system or the probe laser. The probe power was set very low that it did not induce significant thermal effect. In Table 1, we list the best measurement results obtained when an IR radiation of 3.58 μW (calibrated using a thermopile sensor) was incident on the silica microtoroid. At the modulation frequency of 49 Hz, an SNR of 480 was obtained. This corresponds to an NEP of 7.46 nW, or IR intensity of 0.095 mW/cm$^2$. This number is mainly limited by the noise floor and can be improved by using a stabilized probe laser and by isolating the sensor system, for example by placing it in a vacuum environment.

| IR freq (Hz) | Signal (mV) | Std Dev (mV) | Noise (mV) | SNR |
|---|---|---|---|---|
| 10 | 5.765 | 0.0294 | 0.03151 | 182.9578 |
| 49 | 4.982 | 0.0155 | 0.01038 | **479.9615** |
| 244 | 1.817 | 0.0356 | 0.007324 | 248.0885 |
| 1234 | 0.3864 | 0.0117 | 0.01398 | 27.63948 |

Table 1. Response and noise level of the silica microtoroid IR sensor at different IR modulation frequencies. Probe laser power is kept low to minimize heating effect. $CO_2$ laser power was at 1% of its maximum output power.

In conclusion, we developed an uncooled thermal detector for IR radiation using a silica microtoroid resonator, and evaluated its sensitivity and frequency response. Unique thermal and optical properties of the resonator enables us to tune the sensitivity and frequency response of the IR detector. The sensitivity of the IR detector can be improved by isolating it from external acoustic/mechanical noise sources, as well as by optimizing the resonator structure to allow better thermal isolation from the substrate. Owing to its ultrasensitive optical detection scheme, microtoroid IR sensors have the potential to compete with the state of the art microbolometer detectors.

## Acknowledgment

The authors gratefully acknowledge the support from Army Research Office (Grant No. W911NF-11-1-0423).

**Figure 1**

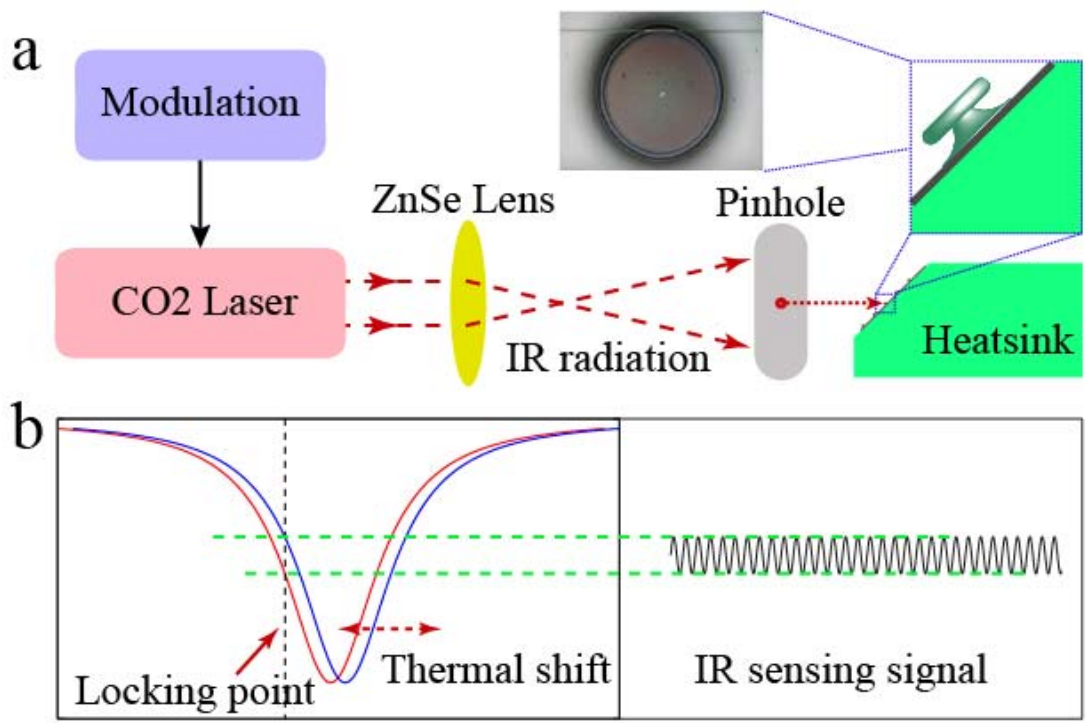

**FIG. 1.** Experimental scheme for characterizing the performance of a microtoroid IR sensor using a $CO_2$ laser. (a) Simplified experimental setup. The inset shows image of microtoroid used in our experiments together with the fiber taper used to couple the probe light in and out of the resonator. (b) Illustration of the IR detection scheme. When the probe laser is locked on the resonance slope, the output signal fluctuates at the modulation frequency of the incoming IR light.

**Figure 2**

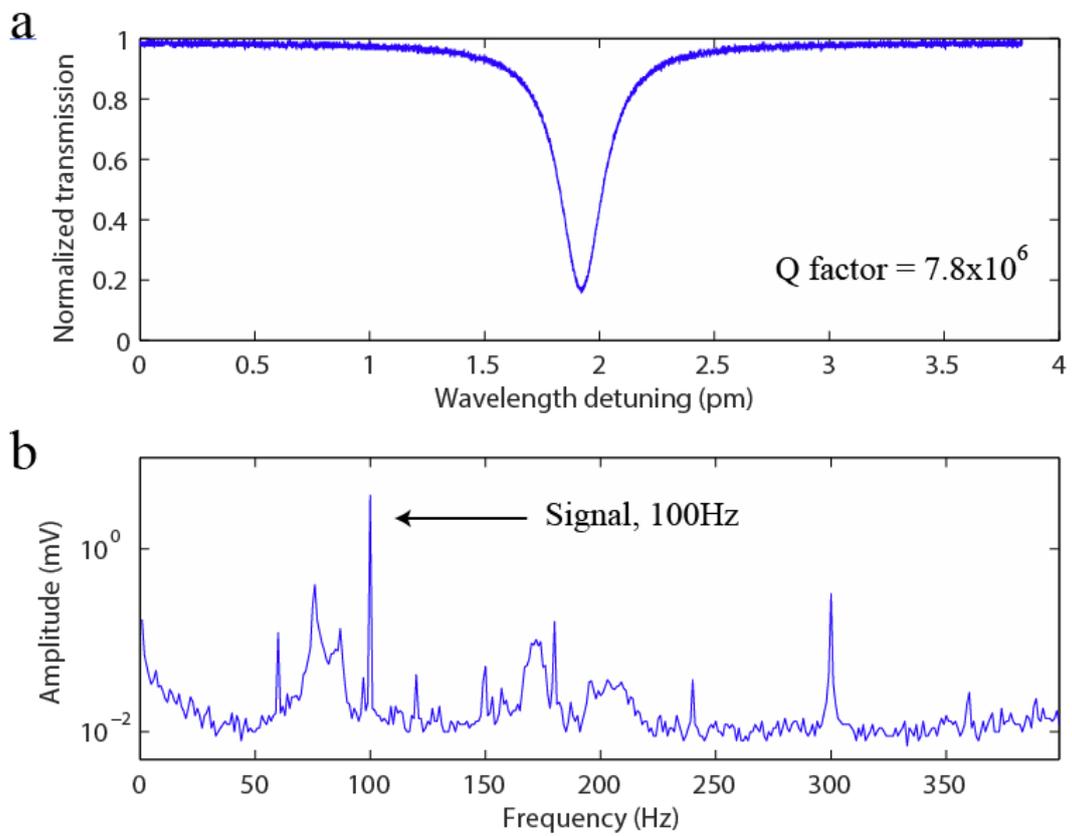

FIG. 2. (a) Transmission spectrum of the microtoroid resonator used to perform the IR measurements. (b) Frequency spectrum of the transmission of the resonator system when the 1450 nm probe laser was locked on the resonance slope with a blue detuning with respect to resonance peak. A clear peak at 100 Hz with SNR of 260 shows the detection of IR radiation. $CO_2$ laser power was set at 5% of the maximum power.

**Figure 3**

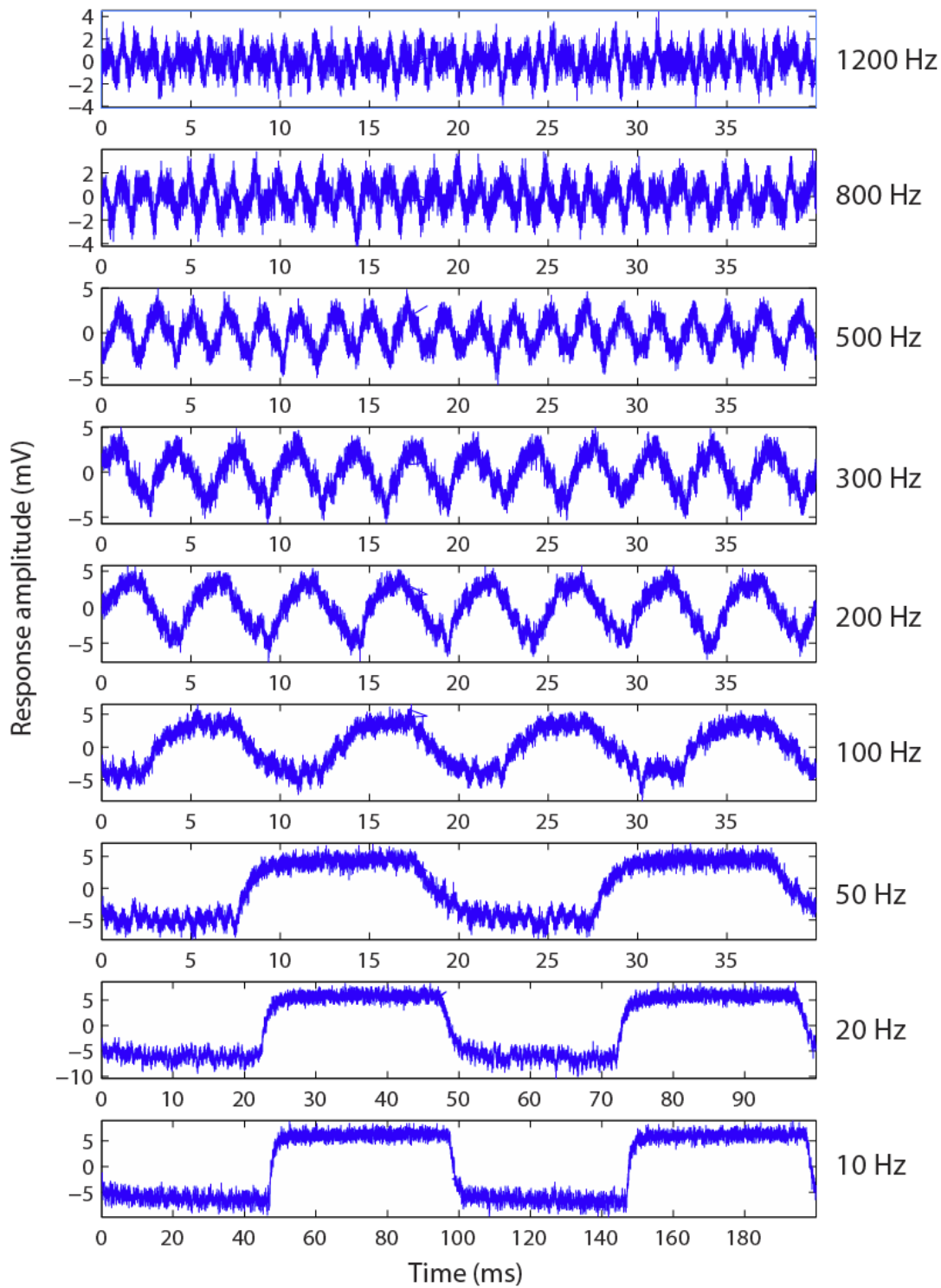

FIG. 3. Response of a microtoroid IR sensor to IR light modulated at various frequencies. When the modulation frequency is low, the transmission response follows the gating pattern of the IR light and displays semi-rectangular wave. At higher modulation frequency, the signal amplitude reduces, and the transmission displays semi-triangle wave. $CO_2$ laser power is set at 10% of the maximum power.

**Figure 4**

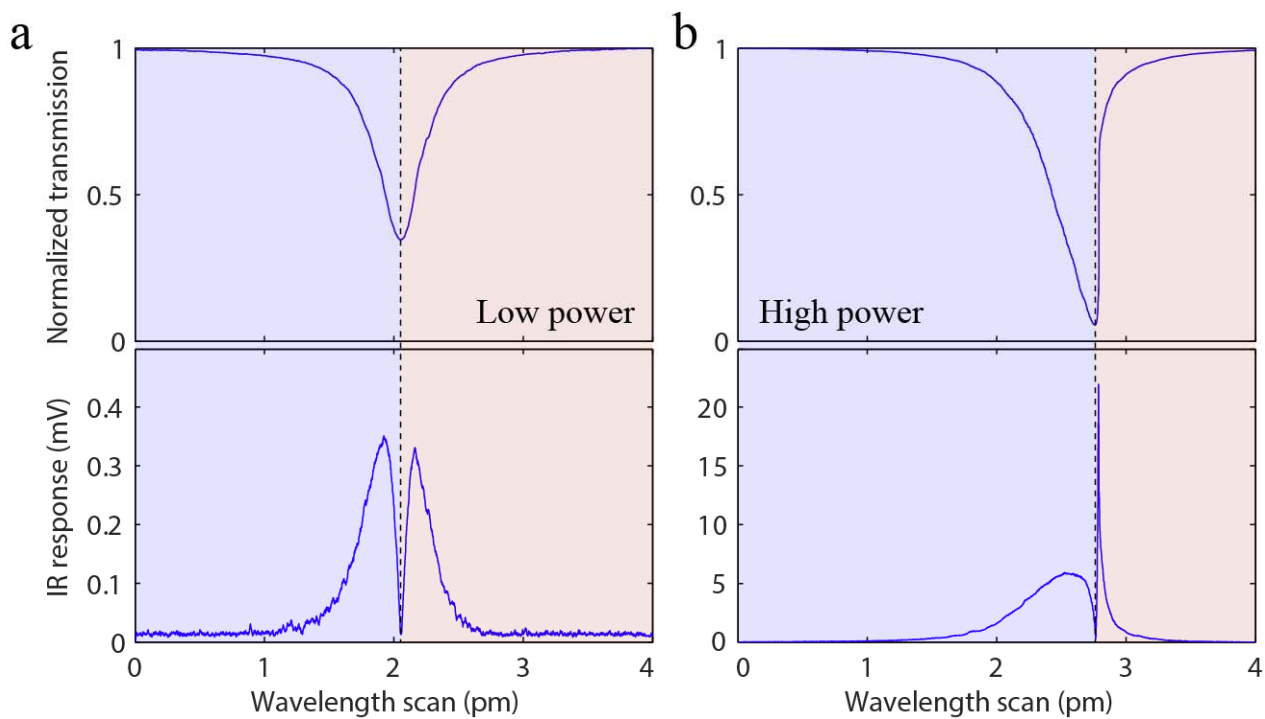

FIG. 4. Measured response of a microtoroid IR sensor at different resonance detuning positions. The probe laser (1437 nm) is linearly scanned for 4 pm in wavelength around the resonance during 500 s scanning time. The $CO_2$ laser used as the IR source was modulated at 25 Hz and its power was set to 10% of its maximum output power. (a) At very lower probe laser power (laser power equivalent to about 10mV), the heating effect of the probe laser is negligible, and the transmission shows a symmetric inverse Lorentzian waveform. IR response is also symmetric on the blue- and red-detuned sides of the resonance. (b) At higher probe laser power (laser power equivalent to 300mV), the heating effect of the probe laser drives the resonance to longer wavelengths and the transmission shows a semi-triangle pattern. The IR response becomes asymmetric with higher response amplitude on the red-detuned side. Each point on the lower panels corresponds to the amplitude of the IR-induced oscillation in the transmission at different frequency-detuning condition.

**Figure 5**

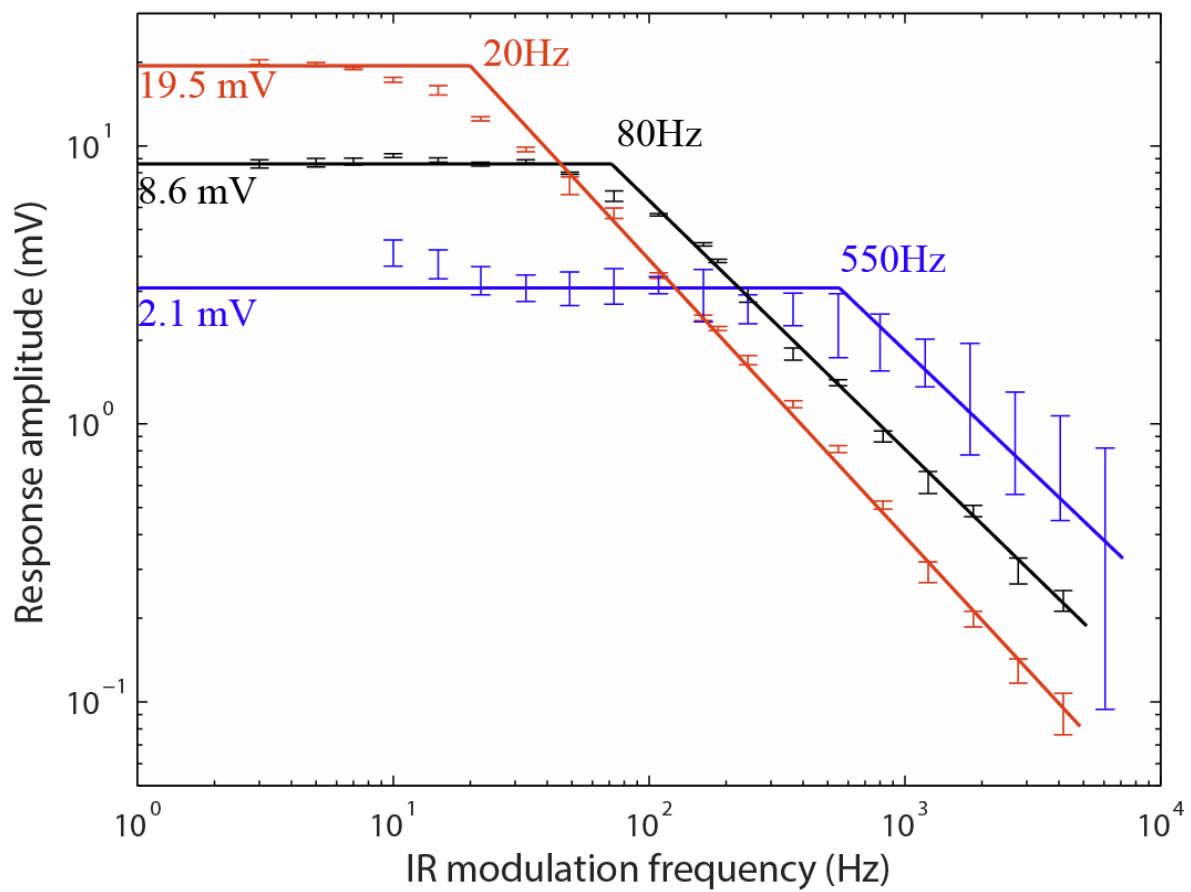

FIG. 5. Frequency response of the microtoroid IR detector at various power and detuning of the probe laser. Red data points were obtained for lower power (laser power equivalent to 330mV) and red-detuned probe laser. Black data points (middle) correspond to lower power (laser power at 330mV) but blue-detuned probe laser. Blue data points correspond to high-power (laser power at 3200mV) and red-detuned probe laser, in which there is a strong heating effect induced by the probe. The $CO_2$ laser power was set at 5% of its maximum output power.